# 01 MODELING VIRTUAL ORGANIZATION ARCHITECTURE WITH THE VIRTUAL ORGANIZATION BREEDING METHODOLOGY

Zbigniew Paszkiewicz, Willy Picard
*Dept. of Information Technology*
*Poznan University of Economics*
*Mansfelda 4, 60-854 Poznan, POLAND*
*{zpasz, picard}@kti.ue.poznan.pl*

*While Enterprise Architecture Modeling (EAM) methodologies become more and more popular, an EAM methodology tailored to the needs of virtual organizations (VO) is still to be developed. Among the most popular EAM methodologies, TOGAF has been chosen as the basis for a new EAM methodology taking into account characteristics of VOs presented in this paper. In this new methodology, referred as Virtual Organization Breeding Methodology (VOBM), concepts developed within the ECOLEAD project, e.g. the concept of Virtual Breeding Environment (VBE) or the VO creation schema, serve as fundamental elements for development of VOBM. VOBM is a generic methodology that should be adapted to a given VBE. VOBM defines the structure of VBE and VO architectures in a service-oriented environment, as well as an architecture development method for virtual organizations (ADM4VO). Finally, a preliminary set of tools and methods for VOBM is given in this paper.*

## 1. INTRODUCTION

The concept of *Virtual Breeding Environment* (VBE) has been proposed by the ECOLEAD project as "an association of organizations and their related supporting institutions, adhering to a base long term cooperation agreement, and adoption of Common operating principles and infrastructures, with the main goal of increasing their preparedness towards collaboration in potential *Virtual Organizations (VO)*" [2]. An important task of each VBE is the support of VO creation [3][4]. While the various steps of the VO creation phase have been studied in the ECOLEAD project [5], a methodology allowing VBE members to model a VO is still to be proposed.

On the other hand, various methodologies addressing enterprise modeling, and more specifically enterprise architecture modeling (EAM), have been proposed in the past 20 years [TOGAF, FEAF, Zachman, MDA] as ways to ensure that the deployed IT infrastructure is aligned with the enterprise's business needs and goals. Methodologies such as TOGAF [11][12][14] and FEAF [15] are well developed, mature and accepted by practitioners.

However these methodologies miss important concepts identified in the ECOLEAD project, e.g. the concepts of VBE, VO composition, partners search and selection, definition of roles in VBE. Therefore, EAM methodologies have to be adapted to the specific characteristics of VBE to be useful for VO modeling

Some works [1][8][9] have studied the benefits that VO modeling would give to VBE members. However, the results presented in formerly mentioned papers were mainly focused on expected benefits without a definition of the methodology allowing VBE members to build the VO models [1] or were not concentrated enough on VBE characteristics [8][9]. Additionally, to our best knowledge, none of these works are based on widely accepted EAM methodologies.

Therefore, there is need for an EAM methodology tailored to the needs for VBEs and VOs. Such a methodology should be built on existing popular EAM methodologies, so that the learning curve of this newly developed methodology would be relatively small for EAM practitioners. The main contribution of this paper is such an EAM methodology, referred as the Virtual Organization Breeding Methodology (VOBM). VOBM is based on the TOGAF methodology, tailored to characteristics of VO and VBE architectures.

This paper is organized as follows. In section 2, enterprise architecture modeling and popular existing methodologies are presented. In 3, the Virtual Organization Breeding Methodology is presented. Then, the Architecture Development Methodology for Virtual Organizations is detailed; Next, VOBM tools and methods are introduced. Finally, section 6 concludes the paper.

## 2. ENTERPRISE ARCHITECTURE MODELING

### 2.1. Overview of enterprise architecture modeling

Enterprise architecture modeling (EAM) aims to provide a formal description of a given system: structure of components, their interrelationships, principles, guidelines, building blocks (applications, people, infrastructure, data etc.), to facilitate system evolution, evaluation, understanding and implementation. In particular, EAM focuses on business goals, strategy, organizational structure, business processes, human and technical resources, competences, information flows [11] Enterprise architecture is typically divided into four architectural domains: business, applications, data and technology [11], each domain having its set of models. This approach and its benefits have been described in [1].
Among EAM methodologies and frameworks the most popular ones [6] are The Open Architecture Group Framework (TOGAF) and the Federal Enterprise Architecture Framework (FEAF).

### 2.2. The Open Group Architecture Framework (TOGAF)

The Open Group Architecture Framework (TOGAF) is a mature and complex methodology for enterprise architecture modeling proposed by the Open Group [13]. The most important three parts in TOGAF are: the Architecture Development

Method (ADM), the TOGAF Enterprise Continuum, and the TOGAF Resource Base. ADM is the core of TOGAF and provides a precise description of all EAM steps leading to an enterprise architecture. The TOGAF Enterprise Continuum is "a "virtual repository" of all the architecture assets - models, patterns, architecture descriptions, and other artifacts - that exist both within the enterprise and in the IT industry at large, which the enterprise considers itself to have available for the development of architectures for the enterprise" [11]. Finally, the TOGAF Resource Base is a set of tools and techniques available for use in applying TOGAF. A detailed description of TOGAF may be found in [11] and [12].

An important aspect of TOGAF is that TOGAF is a generic architecture, i.e. it must be tailored to the need of a specific organization, mainly via the choice of appropriate tools and artifacts. Then, TOGAF is flexible enough to allow architects to use other enterprise frameworks, e.g. the Zachman Framework or Federal Enterprise Architecture Framework (FEAF), within TOGAF.

### 2.3. Federal Enterprise Architecture Framework (FEAF)

The U.S.A Federal CIO Council has defined in FEAF an enterprise architecture for federal agencies or cross-agencies systems. FEAF is based on the concept of *federation of architectures* as "independently developed, maintained and managed architectures that are subsequently integrated within a meta-architecture framework. Such a framework specifies principles for interoperability, migration and conformance. FEAF allows specific business units to have architectures developed and governed as standalone architecture projects." [11] In fact it is not possible to develop organization-wide architectures and keep them integrated, well documented, and flexible. It is necessary to have a number of different architectures existing across an organization and focused on various aspects of the organization.

## 3. OVERVIEW OF THE VIRTUAL ORGANIZATION BREEDING METHODOLOGY

As mentioned in the Introduction, on the one hand, existing EAM methodologies are not tailored to the characteristics of VBEs. On the other hand, an EAM methodology is needed to ease the creation of VOs within the framework presented in the ECOLEAD project.

The solution proposed in this paper is an EAM methodology tailored to VBE and VO creation, referred in this paper as VOBM (Virtual Organization Breeding Methodology). VOBM is based mainly on TOGAF and the results of the ECOLEAD project, but it also takes advantage of concepts present in FEAF.

### 3.1. VOBM as a generic methodology

The aim of the development of VOBM is the definition of a generic methodology providing a set of standard methods and tools for VO modeling. Such a generic methodology then should be customized to the needs and characteristics of a specific VBE. For instance, the characteristics and purpose of a VBE in the construction sector may be significantly distinct to a VBE in the car industry. As a consequence, the shape of created VOs (and therefore their architecture models) will also probably differ from one VBE to another.

In the VOBM approach (*cf.* Figure 1), the VBE manager is responsible for tailoring VOBM to the VBE he/she is managing. Therefore, every VBE should have specific virtual organization architecture modeling framework that would allow creation of virtual organization architectures. In this sense VOBM is a generic methodology. Definition and modeling of VO architecture is a responsibility of VO planners.

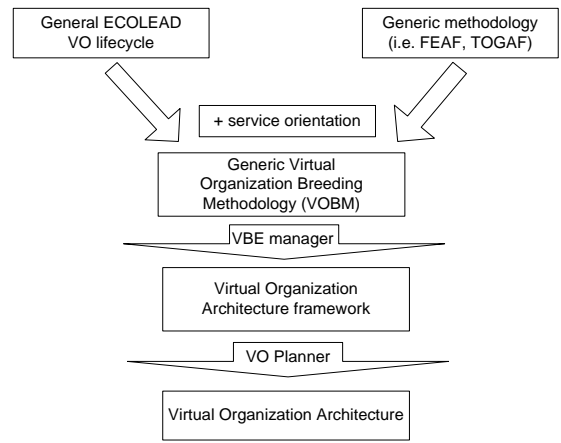

Figure 1: Approach to specification of Virtual Organization Breeding Methodology

### 3.2. Architecture domains in VOBM

TOGAF introduces four architecture domains: business, application, data and infrastructure and two service domains: business and information system services. The *business architecture* defines the business strategy, governance, organization, and key business processes. The *application architecture* provides a blueprint for the individual application systems to be deployed, their interactions, and their relationships to the core business processes of the organization. The *data architecture* describes the structure of an organization's logical and physical data assets and data management resources. Finally, the *technology architecture* describes the logical software and hardware capabilities that are required to support the deployment of business, data, and application services.

Additionally, two levels of services – business services and information system services – are distinguished in TOGAF. Business services are mainly provided at the business architecture domain, while information system services are mainly provided at application, data and infrastructure architecture domains.

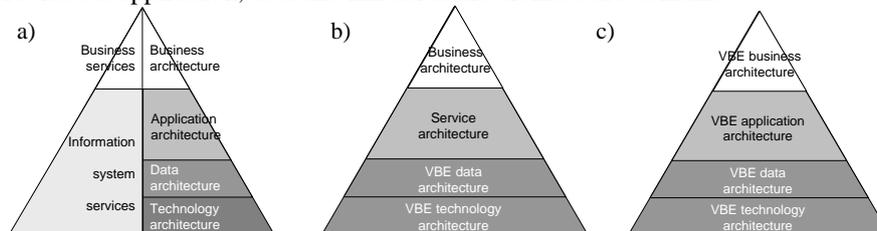

Figure 2: Architecture domains: a) TOGAF b) VO c) VBE

These concepts may be applied to model not only the architecture(s) of organizations being members of the VBE, but also the architecture of the VBE itself. The VBE business architecture domain encompasses e.g. VBE members' role definitions, VBE internal processes. The VBE application architecture domain encompasses applications to be used in all phases of VBE and VO lifecycle, e.g. application supporting the registration of new VBE members or applications for conducting negotiations among potential VO partners. The VBE data architecture domain encompasses, e.g., standard format of documents, definition of data types. The VBE infrastructure architecture domain consists of shared infrastructure and IT solutions or Enterprise Service Bus (ESB) used by VBE for member integration.

VBE architecture domain formation should be a part of VBE instantiation methodology and is not a subject of VOBM. However, VOBM assumes that VBE is organized as formally presented.

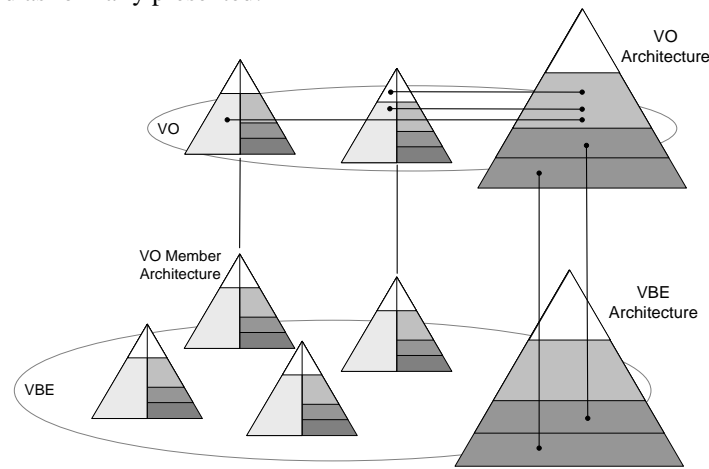

Figure 3 Virtual Organization architecture development. The two lowest architecture domains are shared with the VBE; the service architecture domain integrates services from VO members

The four architecture domain proposed by TOGAF should be adapted to VOs: while the concept of business architecture domain applies to VOs, the three remaining architecture domains should be adapted.

The application architecture domain usually does not apply to VOs, as VOs do not possess their own applications and processes but usually take advantage of applications and processes offered by VO members in a form of services [10]. Ideally, integration of partner services should take place at the highest possible level, i.e. business level. In practice, however, integration at the business level is usually not possible: the operation of a VO usually requires services not only on at the business but also information, data and infrastructure level. Therefore, in the VOBM, the application architecture domain is replaced by a *service architecture* domain which integrates business services and information system services of VO members.

In the VOBM, the data and infrastructure architecture domains of VOs have been tailored to characteristics of VBEs. VBE creates frames for VO creation and operation. These frames include shared standards of data exchanged among VBE members, as well as a common infrastructure. Therefore, in the VOBM, all VOs of a given VBE share the data and infrastructure architecture domains of the VBE.

The architecture domains of VBE members (.i.e. TOGAF architecture domains), VBEs (as defined in VOBM), and VOs (as defined in VOBM) are illustrated in Figure 2. The composition of a VO architecture in the VOBM is illustrated in Figure 3.

## 4. ADM FOR VIRTUAL ORGANIZATIONS

A core element of the VOBM is the Architecture Development Method for Virtual Organizations (ADM4VO). ADM4VO is a tailored version of the Architecture Development Method (ADM) developed in TOGAF, adapted to the characteristics of VOs. In ADM4VO, a number of ADM components have been reviewed, modified, added or removed. While ADM consists of 10 phases, ADM4VO consists of only 6 phases as the following phases of TOGAF ADM have been left out:

- Preliminary phase: in ADM, this phase includes the choice of tools and methods to be used. In ADM4VO, these decisions are made by the VBE manager during the adaptation of VOBM to VBE needs.
- Information Systems Architecture and Technology Architecture phases: these phase have been replaced by the VO Service Architecture phase, as VOs do not possess their own systems and infrastructure, but compose processes and system using services provided by VO members.
- Migration planning and Implementation governance phases: these phases are strictly connected with the implementation of the modeled architecture, i.e. development and deployment of applications and infrastructure. Some aspects of this phase are included in VO Service Architecture phase.

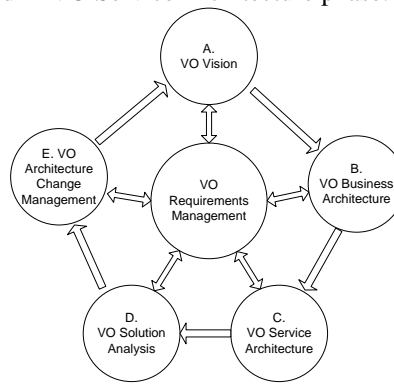

Figure 4. Architecture Development Method for Virtual Organization (ADM4VO)

Two phases of ADM4VO – VO Architecture Change Management and Requirement Management – do not have an equivalent in ECOLEAD VO creation methodology, while existing in ADM. VO Architecture Change Management focuses on the adaptation of VO to changing environment/circumstances, on conditions under which the architecture, or parts of it, will be permitted to change. The VO Architecture Change Management phase should exploit advanced concepts of business process adoption, change and impact analysis.

The Requirements Management phase focuses on monitoring every VO architecture creation step and verifying the results of these steps against requirements, assumptions and results of the previous phases. For instance, if a set of services is identified during the Service Architecture phase as needed by the VO, but it appears as a result of the VO Solution Analysis phase that these services are not provided by any member of the VBE, then the Architecture Service assumptions have to be verified and changed, which may lead to changes in Business Architecture or even VO Vision. The Requirements Management phase provides a mechanism for back-loops within one architecture development cycle.

The goals of other phases are listed below:

**Phase A: VO Architecture Vision**
- Definition of relevant stakeholders, their concerns and objectives, key business requirements;
- Articulation of goals and requirements for VO and VO architecture.

**Phase B: VO Business Architecture**
- Development of a Target Business Architecture, describing the organizational, functional, process and information aspects of the business environment;
- Description of the Baseline Business Architecture and gap analysis between the Baseline and Target Business Architectures when a VO already exists and the aim of ADM4VO is to adopt it to changing environment.

Modeling can go far beyond simple business process modeling, taking advantage of already existing building blocks present in Architecture Repository (VBE bag of assets).. This phase includes also definition of VO requirements (needed competences and partners) on a basis of business process.

**Phase C: VO Service Architecture**
- Development of a Target Service Architecture describing the needed services, and competences of partners;
- Description of the Baseline Service Architecture and gap analysis between the Baseline and Target Service Architectures when a VO already exist and the aim of ADM4VO is to adopt it to changing environment,
- Choice of partners for cooperation, based on competencies, service description, and on information about social network

This phase does not exist in TOGAF ADM. It has been developed for the need of VOBM. Phase C is an initial partner selection phase. The result of this phase is the composition of services offered by selected partners into complex business models developed in phase B.

**Phase D: VO Solution Analysis**
- Evaluation in terms of costs, benefits, potential problems etc. and selection among various possible implementations of business and service architectures,

The aim of this phase in ADM (very briefly described in TOGAF) is to evaluate in terms of costs and benefits various possibilities of organization architecture implementation. In ADM4VO, the idea of comparing many possible variants of VO (variants differ by selected partners, services, etc.) is similar, even if the concept of service plays a greater role than in ADM.

## 5. VOBM TOOLS AND METHODS

An important element of each EAM methodology is the identification of methods and tools to be used in each phase. In table 1, a preliminary list of methods and tools has been established. It should be noted that this list is not exhaustive. For phases common for ADM and ADM4VO, methods and tools defined in TOGAF can be used. Similarly, tools and methods identified by the ECOLEAD project should be evaluated as regards VOBM. For phases specific to ADM4VO, e.g. the VO Service Architecture phase, new tools and methods still need to be identified and developed.

Table 1 Virtual Organization Breeding Methodology steps and methods outline

|     | VOBM phases | Methods |
| --- | --- | --- |
| A   | VO Vision | ‣ Business scenario method<br>‣ Methods for aligning goals with business process, partners and services for change and impact analysis |
| B.1 | VO Business Architecture<br>*VO Business Modeling* | ‣ VBE bag of assets and building block concepts<br>‣ Multilayered cooperation protocols<br>‣ Methods for protocol selection<br>‣ Protocol to business process mapping |
| B.2 | VO Business Architecture<br>*VO Requirements Definition* | ‣ Roles definition<br>‣ Methods, models and languages for skill requirements specification |
| C   | VO Service Architecture | ‣ Partner selection methods<br>‣ VO ontology<br>‣ Service selection methods<br>‣ Social networks and social protocols concepts<br>‣ Composition of services into complex business process |
| D   | VO Solution Analysis | ‣ Methods for service management, evaluation and verification<br>‣ Tools for negotiation<br>‣ Key Performance Indicators for Virtual Organization |
| E   | VO Change Management | ‣ VO and business process adaptation<br>‣ Change and impact analysis methods |
| F   | VO Requirements Management | ‣ Change and impact analysis methods<br>‣ Key Performance Indicators for Virtual Organization<br>‣ VO requirements management methods |

## 6. CONCLUSIONS

The main contribution presented in this paper is a new EAM methodology – the Virtual Organization Breeding Methodology (VOBM) – which is tailored to the characteristics of VOs and concepts identified by the ECOLEAD project. VOBM is based on one of the most popular EAM methodologies, TOGAF, which ensures a relatively small learning curve for TOGAF practitioners.

As regards TOGAF, VOBM introduces Service Architecture in VO architecture domains, the VBE architecture, and modified development phases. As regards ECOLEAD, VOBM introduces Change and Requirements Management phases to support VO agility. Introduced changes are oriented to use SOA approach.

Among future works, a detailed specification of each phase of ADM4VO is still to be written. Additionally, both TOGAF and ECOLEAD provide a set of tools, artifact templates, and methods which should be evaluated as regards their potential use in VOBM.

Finally, within the context of the IT-SOA project [7], a service-oriented VBE is currently under development. This VBE will gather companies from the construction sector in the Great Poland region. During the IT-SOA project, the VOBM will be evaluated in a real business environment.